\documentclass[12pt]{article}
\usepackage{color}
\usepackage[dvips]{graphicx}

\begin{document}
\begin{center}
\Large{
Theoretical Study of Reactivity Indices and Rough Potential Energy Curves
for the Dissociation of 59 Fullerendiols in Gas-Phase and in
Aqueous Solution with an Implicit Solvent Model} 
\end{center}

\noindent Anne Justine ETINDELE
\textit{
Higher Teachers Training College, University of Yaounde I,
P.O. Box 47, Yaounde, CAMEROON\\
e-mail: annetindele@yahoo.fr\\
}

\noindent Abraham PONRA\\
\textit{
{Department of Physics, Faculty of Science, University of Maroua, 
P.O.\ Box 814, Maroua, CAMEROON\\
e-mail: abraponra@yahoo.com\\
}}


\noindent Mark E. CASIDA\\
\textit{
Laboratoire de Spectrom\'etrie, Interactions et Chimie th\'eorique (SITh), 
D\'epartement de Chimie Mol\'eculaire (DCM, UMR CNRS/UGA 5250), 
Institut de Chimie Mol\'eculaire de Grenoble (ICMG,
FR2607), Universit\'e Grenoble Alpes (UGA) 301 rue de la Chimie, BP 53, F-38041 Grenoble Cedex
9, FRANCE\\
e-mail: mark.casida@univ-grenoble-alpes.fr\\
}

\noindent G.\ Andr\'es CISNEROS\\
\textit{
Department of Physics, Department of Chemistry and Biochemistry,
University of Texas at Dallas,
Richardson, Texas, 75802, USA\\
e-mail: andres@utdallas.edu\\
}

\noindent Jorge NOCHEBUENA\\
\textit{
Department of Chemistry and Biochemistry,
College of Science and Mathematics,
Augusta University,
Augusta, Georgia, 30912, USA\\
e-mail: jnochebuenaherna@augusta.edu\\
}

\vspace{0.5cm}

%

\begin{center}
{\bf Abstract}
\end{center}
Buckminsterfullerene, C$_{60}$, has not only a beautiful truncated
icosahedral (soccerball) shape, but simple H\"uckel calculations
predict a three-fold degenerate lowest unoccupied molecular orbital
(LUMO) which can accomodate up to six electrons making it a good
electron acceptor.  Experiments have confirmed that C$_{60}$ is a
radical sponge and it is now sold for use in topical cosmetics.
Further medical uses require functionalization of C$_{60}$ to make
it soluble and one of the simplest functionalization is to make
C$_{60}$(OH)$_n$ fullerenols.  A previous article [{\em Adv.\ Quant.\ Chem.}\
{\bf 8}, 351 (2023)] studied reactivity indices for the successive 
addition of the $^\bullet$OH radical to ($^\bullet$)C$_{60}$(OH)$_n$ 
in gas phase.  [($^\bullet$)C$_{60}$(OH)$_n$ is only a radical when
$n$ is an odd number.] This present article extends this previous
work by examining various aspects of how the reaction, 
\begin{displaymath}
  \mbox{$^\bullet$C$_{60}$OH} + \mbox{$^\bullet$OH} \rightarrow
  \mbox{C$_{60}$(OH)$_2$} \, \, \, (1) \nonumber 
\end{displaymath}
changes in aqueous solution.  One obvious difference between 
C$_{60}$ and their various isomers of C$_{60}$(OH)$_2$ is the presence
of a dipole.  As fullerendiols are nearly spherical, their change
in dipole moment in going from gas to aqueous phase may be estimated
using back-of-the-envellope calculations with the Onsager model.  The
result is remarkably similar to what is obtained using density-functional
theory (DFT) witn an implicit solvation model (Surface Molecular Density, SMD).  Calculation of fullerendiol C-O bond 
energies and reactivity indices using the SMD approach confirm that the 
general conclusions from the earlier work regarding gas-phase reactivity 
still hold in the aqueous phase.  A major difference between the present 
work and the earlier work is the calculation of potential
energy curves (PECs) for reaction (1) in gas and aqueous phases.  This is
done in exploratory work for all 59 possible fullerendiols in both gas
phase and in aqueous solution with the SMD approach using spin-unrestricted 
DFT calculations with symmetry breaking.  Surprisingly little change is
found between the gas-phase and aqueous-phase PECs.  It is discovered
that the majority of C$_{60}$(OH)$_2$ show radicaloid character, as might
have been expected from trying to draw resonance structures.  
Spin-contamination curves are also remarkably similar for
gas-phase and aqueous-phase results.  Although our calculations do not 
include a dispersion correction, it was noticed that all calculated PECs
have a $1/R^6$ behavior over a significant $R = R(\mbox{C-O})$ distance, 
underlying the need to be careful of double counting when including 
dispersion corrections in DFT.  A short coming of
our SMD approach is the lack of explicit water molecules which can form
hydrogen bonds with the OH groups and dissociating radicals.  


\newpage
\section{Introduction}
\label{sec:intro}

\marginpar{\color{blue} C$_{60}$}
Buckminsterfullerene C$_{60}$ is well-known for its geometrical beauty, 
reminiscent of the geodesic domes of Buckminsterfuller. It can also capture 
\marginpar{\color{blue} LUMO}
up to six electrons in its $t_{1u}$ lowest unoccupied molecular orbitals 
(LUMOs). This strongly electrophilic property has been characterized 
experimentally and it has been called a ``radical sponge'' \cite{MML92}. 
It is even used as a commercial ingredient in some skin care products, 
one of which is an anti-aging moisturizer which goes by the name ``C60'' 
\cite{C60skincream}. However many applications require an increase in the 
aqueous solubility of this hydrophobic molecule. A particularly simple way 
to increase the solubility of C$_{60}$ is to decorate it with hydroxyl groups. 
As C$_{60}$ is a ``radical sponge,'' it can react with a large number of 
hydroxyl radicals 
\cite{CMW+00,RG04,KMT+08,FR09,RTS+11,UYA+14,SCP+16,AKMM17,KT17,VGG18,%
WGZ18,KSV+19,ZSM+20,PEMC23}
to create fullerenols ($^\bullet$)C$_{60}$(OH)$_n$ where the bullet 
($^\bullet$) is a reminder that these fullerenerols are radicals for odd $n$.

To take into account the impact of solvents on the properties of molecules, 
two approaches may be considered. The first approach takes the molecular nature
of the solvent molecules explicitly into account. 
We will only use the second in the present work.  This is 
the implicit model in which the solvent is considered 
as a dielectric continuum.
As the ($^\bullet$)C$_{60}$(OH)$_n$ are nearly spherical, Onsager's 
 spherical dielectric cavity reaction field model provides 
a pencil and paper way to determine the dipole moment for the molecule in 
solution from its gas-phase dipole moment.
Thus Onsager’s spherical dielectric cavity reaction field model provides 
a pencil and paper way to determine the dipole moment for the molecule in 
solution from its gas-phase dipole moment.   
This first-order approximation works almost surprisingly well compared to
other implicit solvent models that use cavities which are more 
carefully adjusted to reflect the shape of the molecule.

The global and local reactivity indices can be significantly impacted 
depending on whether the study is carried out in the gas or aqueous phase
\cite{MCC+24,FS19}. These reactivity indices made it possible to effectively 
study the donor and acceptor character of C$_{60}$(OH)$_n$ \cite{PEMC23} 
in the gas phase. So, what about this character for the aqueous phase? 
We are given the opportunity to try to answer this question in this paper. 
\marginpar{\color{blue} PEC}
In addition, the description of potential energy curves (PECs) could 
provide valuable information on the impact of the presence of the solvent 
on the dissociation (or formation) of the fullenerol-hydroxyl 
(i.e., $^\bullet$C$_{60}$(OH) + $^\bullet$OH) bonds. 

Previous work has shown that the dissociation of a molecule into two 
fragments requires taking into account symmetry breaking along the 
PECs \cite{PEMC21,PBEC23}.  Taking into account this 
symmetry breaking, point-by-point, throughout the PECs provides a unique 
opportunity to determine the Coulson-Fisher points for the 59 isomers of 
fullerenediol by examining the variation of $\langle {\hat S}^2 \rangle$
with distance.

This article is organized in the following manner: The next section covers 
the basic theory used in the rest of the paper. Section~\ref{sec:details} 
gives the computational details that we are using in this paper for all 
the calculations. Section~\ref{sec:results} presents and discusses our 
results.  In particular, we present PECs and their associated
$\langle {\hat S}^2 \rangle$ for all 59 isomers in the SI. 
Section~\ref{sec:conclude} is the concluding discussion.

%
%


\section{Theory}
\label{sec:theory}

Electronic structure calculations were initially done for atoms, gradually
moved onto small molecules, and then---with the development of band 
theory---onto crystals.  However it is important to be able to model the
electronic structure of solvated molecules for the simple reason that
most experiments are carried out in solution, or at least involve key
steps and/or measurements carried out in solution.  We use 
an implicit solvent model.  This replaces the solvent with a dielectric 
cavity.  It has the advantage of being relatively simple and so 
computationally expedient, but does not take into account all of the
physics of a molecule in solution.  


\marginpar{\color{blue} SMD}
The implicit solvent model used in the present work is the well-known
SMD (for Solvation Model based upon the quantum mechanical Density) 
\cite{MCT09}.  This is a reaction field model where the solvent molecules
are replaced by a dielectric continuum surrounding the solvated molecules
\marginpar{\color{blue} SEC}
in a solvent excluded cavity (SEC, also said to be defined by a solvent 
accessible surface).  The molecule in the cavity is described by 
\marginpar{\color{blue} QM}
quantum mechanics (QM), from which its charge distribution is calculated.
This charge distribution creates a reaction field in the surrounding
dielectric which, in turn, modifies the QM calculation.  Although this
sounds like it requires an iterative self-consistent calculation, its
solution may be done in a non-iterative single-step procedure.  Particular
implicit solvent models vary by how the SEC is generated and how the Poisson
equation is solved to determine the reaction field.  In particular, the 
generation of the SEC involves some semi-empirical parameters which are fit,
in the SMD model, to experimental solvation free energies.  It is not our 
purpose to go further into details here.

However fullerenediols are nearly spherical molecules which means that
the oldest most approximate reaction field model---namely the Onsager
model \cite{O36} is expected to be a good approximation.  Although
a program such as {\sc Gaussian} \cite{gaussian} has an option to do
Onsager reaction field calculations, it is really a shame to use such
a large program for what is ultimately a back-of-the-envelope calculation.
Indeed, doing the back-of-the-envelope calculation leads to enhanced
understanding and so we do present this model here.  As with more sophisticated 
implicit solvent models, the solvent is treated as a polarizable
continuum with dielectric constant $\epsilon$.  However the molecule 
is treated as a point dipole moment $\mu_{\ell}$ whose value will depend 
upon its surroundings.  It also has a polarizability ${\bf \alpha}$.  
The molecule is placed in a solvent-excluded spherical cavity of radius $a$.  
The electric field
of the molecular dipole moment will cause a redistribution of charges
around the sphere and induce a reaction (electric field) ${\vec {\cal E}}$
inside the sphere which acts on the molecule to change the dipole moment
through the equation,
\begin{equation}
  {\vec \mu}_{\ell} = {\vec \mu}_g + {\bf \alpha} {\vec {\cal E}} \, ,
  \label{eq:Onsager.1}
\end{equation}
where ${\vec \mu}_{\ell}$ is the molecular dipole moment in the liquid and
$\mu_g$ is the gas-phase dipole moment.  This reaction field is
(Ref.~\cite{B73} pp.~130-134)
\begin{equation}
  {\vec {\cal E}} = \frac{2(\epsilon-1)}{2 \epsilon+1} \frac{1}{a^3} {\vec \mu}_{\ell}
  \, .
  \label{eq:Onsager.2}
\end{equation}
Therefore,
\begin{equation}
  {\vec {\cal E}} = \frac{2(\epsilon-1)}{2 \epsilon+1} \frac{1}{a^3} \left(
  {\vec \mu}_g + {\bf \alpha} {\vec {\cal E}} \right)
  \, ,
  \label{eq:Onsager.3}
\end{equation}
which can be solved for the reaction field to give,
\begin{equation}
  {\vec {\cal E}} = \left( \frac{2\epsilon+1}{2(\epsilon-1)} a^2 {\bf 1} - {\bf \alpha}
   \right)^{-1} {\vec \mu}_g \, .
   \label{eq:Onsager.4}
\end{equation}
Finally, plugging this result into Eq.~(\ref{eq:Onsager.1}) and rearranging
gives,
\begin{equation}
  {\vec \mu}_{\ell} = \left( {\bf 1} - \frac{2(\epsilon-1)}{2\epsilon+1}
  \frac{1}{a^3} {\bf \alpha} \right)^{-1} {\vec \mu_g} \, .
  \label{eq:Onsager.5}
\end{equation}
This still leaves the problem of the polarizability and the size of the
cavity.  However assuming an isotropic medium, so that we may replace the
polarizability tensor ${\bf \alpha}$ with the average polarizability
${\bar \alpha}$, gives,
\begin{equation}
  {\vec \mu}_{\ell} = \left( 1 - \frac{2(\epsilon-1)}{2\epsilon+1}
  \frac{\bar \alpha}{a^3} \right)^{-1} {\vec \mu_g} \, ,
  \label{eq:Onsager.6}
\end{equation}
so that we actually only need to determine ${\bar \alpha}/a^3$.  
This can be done using the Lorenz-Lorentz equation relating the polarizabilty 
and the refractive index $n$,
\begin{equation}
  \frac{n^2-1}{n^2+2} = \frac{4\pi}{3} \rho {\bar \alpha} \, ,
  \label{eq:Onsager.7}
\end{equation}
where $\rho$ is the number density so that,
\begin{equation}
  \frac{1}{\rho} = \frac{4}{3} \pi a^3 \, .
  \label{eq:Onsager.8}
\end{equation}
Hence,
\begin{equation}
  \frac{\bar \alpha}{a^3} = \frac{n^2-1}{n^2+2} \, .
  \label{eq:Onsager.9}
\end{equation}
Note that using these equations for treating C$_{60}$ in water requires
the dielectric constant for water but the refractive index of C$_{60}$.
For water, $\epsilon=80$ at 20 $^\circ$C (Ref.~\cite{P63}, p.~299).
The refractive index of solid C$_{60}$ which is 2.2 at 630 nm wavelength
\cite{H91}.  Putting everything together, then
\begin{equation}
  {\vec \mu}_{\ell} = 2.23 {\vec \mu}_g \, .
  \label{eq:Onsager.10}
\end{equation}
for C$_{60}$ in aqueous solution.  We expect similar results for fullerenols.

\section{Computational Details}
\label{sec:details}



\begin{figure}
\begin{center}
\includegraphics[width=0.4\textwidth]{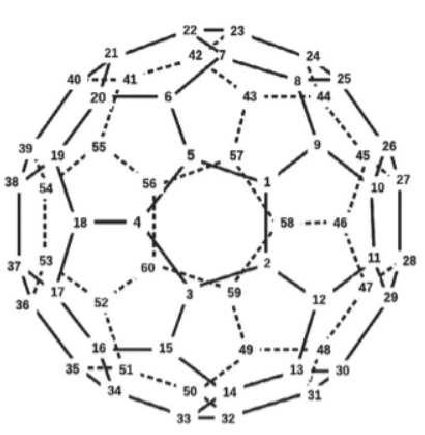}
\end{center}
\caption{
IUPAC numbering of buckminsterfullerene \cite{PCM+02}.
\label{fig:IUPAC}
}
\end{figure}

\marginpar{\color{blue} IUPAC}
The International Union of Pure and Applied Chemistry (IUPAC) 
numbering ({\bf Fig.~\ref{fig:IUPAC}}) is used throughout for 
C$_{60}$ \cite{PCM+02}. An ``xyz'' format file with this numbering has 
\marginpar{\color{blue} ESI}
been supplied in the Electronic Supplementary Information (ESI) 
for use with molecular visualization software.


{\em 
The same level of electronic structure calculation is used throughout
this article, whether it be for the gas-phase or for the aqueous phase
using an implicit 
solvent model.
}

Calculations were carried out using version 5.0.4 of 
{\sc Orca} \cite{N18} and
with {\sc Gaussian} \cite{gaussian}.  The B3LYP keyword was
used with {\sc Gaussian} while the B3LYP/g keyword was used with
{\sc Orca}.  This is the same functional which we will rename
B3LYP(VWN3) \cite{B93,LYP88} because the original B3LYP functional (programmed
in {\sc Gaussian}) used the Vosko-Wilk-Nusair parameterization of the
random-phase approximation correlation energy (known as VWN3 in 
{\sc Gaussian}) but that the default B3LYP functional in {\sc Orca}
uses the Vosko-Wilk-Nusair parameterization of the Ceperley and Alder's
quantum Monte Carlo correlation energy (known as VWN5 in {\sc Gaussian})
unless ``/g'' is specified.  The same def2SVP \cite{WA05} orbital basis set
was used in all our calculations.

\marginpar{\color{blue} RI}
The default in {\sc Orca} is also to use a resolution-of-the-identity (RI)
approximation to reduce the problem of calculating four-center electron
repulsion integrals to that of calculating only three-center electron
repulsion integrals.  As this option was used in our previous gas-phase 
work \cite{PEMC23} on gas-phase reactivity indices, we continued with 
the same option in the article when calculating reactivity indices in 
solution.  It was also used in calculating gas and solution phase dipole
moments.  However it was judged useful to follow the advice of 
Ref.~\cite{MDC24} in order to ensure the best possible agreement
between {\sc Orca} and {\sc Gaussian} calculations:
\begin{quote}
\noindent
`` In ORCA, keywords used included {\tt NORI} (no approximation is used), 
{\tt TightSCF}, {\tt TightOpt}, {\tt SlowConv}, {\tt NumFreq} and an 
ultra-fine grid. We explicitly confirmed that {\sc Gaussian} and 
{\sc Orca} gave the same results for the same basis sets and functionals 
in single-point calculations.'' \cite{MDC24}
\end{quote}

Implicit solvent calculations were carried out with 
{\sc Orca} \cite{N18} at the B3LYP(VWN3)/def2-SVP level using the 
SMD \cite{MCT09} solvent model.  It has been discovered,
in work on polypyridine ruthenium clusters, that {\sc Gaussian} and 
{\sc Orca} use different grids on the solvent-allowed 
surface which leads to significant differences in numerical results  
(see the ESI of Ref.~\cite{DMC24}).  Following that work, we opted to use
\begin{verbatim}
%cpcm
num_leb 590
end
\end{verbatim}
in the {\sc Orca} input to in order ensure good agreement with 
{\sc Gaussian}.

\section{Results}
\label{sec:results}

\begin{table}
\begin{center}
\begin{tabular}{ccc}
\hline \hline
Count & Symmetry-Equivalent Isomers  & Distance \\
\hline
2       & (1,2)-C$_{60}$(OH)$_2$, (1,5)-C$_{60}$(OH)$_2$   & 1 bond \\
+2 = 4  & (1,3)-C$_{60}$(OH)$_2$, (1,4)-C$_{60}$(OH)$_2$   & 2 bonds \\ 
+2 = 6  & (1,6)-C$_{60}$(OH)$_2$, (1,12)-C$_{60}$(OH)$_2$  & 2 bonds \\
+2 = 8  & (1,7)$^*$-C$_{60}$(OH)$_2$, (1,11)$^*$-C$_{60}$(OH)$_2$  & 3 bonds \\
+2 = 10 & (1,8)$^*$-C$_{60}$(OH)$_2$, (1,10)$^*$-C$_{60}$(OH)$_2$  & 2 bonds \\ 
+1 = 11 & (1,9)-C$_{60}$(OH)$_2$                           & 1 bond \\
+2 = 13 & (1,13)-C$_{60}$(OH)$_2$, (1,20)-C$_{60}$(OH)$_2$ & 3 bonds \\
+2 = 15 & (1,14)-C$_{60}$(OH)$_2$, (1,19)-C$_{60}$(OH)$_2$ & 4 bonds \\
+2 = 17 & (1,15)$^*$-C$_{60}$(OH)$_2$, (1,18)$^*$-C$_{60}$(OH)$_2$ & 3 bonds \\
+2 = 19 & (1,16)-C$_{60}$(OH)$_2$, (1,17)-C$_{60}$(OH)$_2$ & 4 bonds \\
+2 = 21 & (1,21)-C$_{60}$(OH)$_2$, (1,30)-C$_{60}$(OH)$_2$ & 4 bonds \\
+2 = 23 & (1,22)$^*$-C$_{60}$(OH)$_2$, (1,29)$^*$-C$_{60}$(OH)$_2$ & 4 bonds \\
+2 = 25 & (1,23)-C$_{60}$(OH)$_2$, (1,28)-C$_{60}$(OH)$_2$ & 5 bonds \\
+2 = 27 & (1,24)$^*$-C$_{60}$(OH)$_2$, (1,27)$^*$-C$_{60}$(OH)$_2$ & 4 bonds \\
+2 = 29 & (1,25)-C$_{60}$(OH)$_2$, (1,26)-C$_{60}$(OH)$_2$ & 3 bonds \\
+2 = 31 & (1,31)-C$_{60}$(OH)$_2$, (1,40)-C$_{60}$(OH)$_2$ & 5 bonds \\
+2 = 33 & (1,32)$^*$-C$_{60}$(OH)$_2$, (1,39)$^*$-C$_{60}$(OH)$_2$ & 6 bonds \\
+2 = 35 & (1,33)$^*$-C$_{60}$(OH)$_2$, (1,38)$^*$-C$_{60}$(OH)$_2$ & 5 bonds \\
+2 = 37 & (1,34)$^*$-C$_{60}$(OH)$_2$, (1,37)$^*$-C$_{60}$(OH)$_2$ & 5 bonds \\
+2 = 39 & (1,35)-C$_{60}$(OH)$_2$, (1,36)-C$_{60}$(OH)$_2$ & 6 bonds \\
+2 = 41 & (1,41)-C$_{60}$(OH)$_2$, (1,48)-C$_{60}$(OH)$_2$ & 6 bonds \\
+2 = 43 & (1,42)-C$_{60}$(OH)$_2$, (1,47)-C$_{60}$(OH)$_2$ & 6 bonds \\
+2 = 45 & (1,43)-C$_{60}$(OH)$_2$, (1,46)-C$_{60}$(OH)$_2$ & 6 bonds \\
+2 = 47 & (1,44)-C$_{60}$(OH)$_2$, (1,45)-C$_{60}$(OH)$_2$ & 5 bonds \\
+2 = 49 & (1,49)-C$_{60}$(OH)$_2$, (1,55)-C$_{60}$(OH)$_2$ & 7 bonds \\
+2 = 51 & (1,50)-C$_{60}$(OH)$_2$, (1,54)-C$_{60}$(OH)$_2$ & 7 bonds \\
+2 = 53 & (1,51)-C$_{60}$(OH)$_2$, (1,53)-C$_{60}$(OH)$_2$ & 6 bonds \\
+1 = 54 & (1,52)-C$_{60}$(OH)$_2$                          & 8 bonds \\
+2 = 56 & (1,56)$^*$-C$_{60}$(OH)$_2$, (1,59)$^*$-C$_{60}$(OH)$_2$ & 8 bonds \\
+2 = 58 & (1,57)-C$_{60}$(OH)$_2$, (1,58)-C$_{60}$(OH)$_2$ & 7 bonds \\
+1 = 59 & (1,60)-C$_{60}$(OH)$_2$                          & 9 bonds \\
\hline \hline
\end{tabular}
\end{center}
\caption{\label{tab:isomercount} List of symmetry equivalent (mirror image) 
isomers.  The distance is the shortest through-bond distance.  Isomers
with an asterisk (*) are symmetry pairs with significant disagreement in their
energy at $R$(C-O) = 1.4 {\AA} (see text).}
\end{table}
A thorough study of the gas-phase reactivity of fullerenols with respect
to step-wise addition of hydroxyl radicals has already been presented
in Ref.~\cite{PEMC23}. Our interest here is in how the physical and
chemical properties of fullerenols differ in gas phase and in 
aqueous solution.  We will confine our study to fullerenediols
only.  As {\bf Table~\ref{tab:isomercount}} shows, this is still a very
large number of isomers.  We will begin with physical properties and
then chemical reactivity indices before going on to 
\marginpar{\color{blue} PEC}
potential energy curves (PECs) for the dissociation reaction,  
\begin{equation}
  \mbox{C$_{60}$(OH)$_2$} \rightarrow \mbox{$^\bullet$C$_{60}$OH}
  + \mbox{$^\bullet$OH} \, .
  \label{eq:results.1}
\end{equation}

\begin{figure}
\begin{center}
\includegraphics[width=0.8\textwidth]{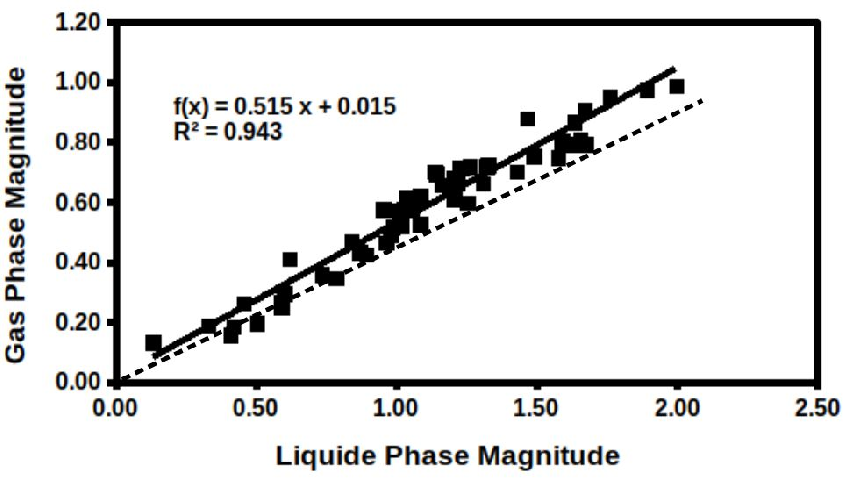}
\end{center}
\caption{
Graph correlating the magnitudes of gas-phase and SMD aqueous-phase
dipole moments for C$_{60}$(OH)$_2$ fullerenols calculated using 
geometries optimized in the gas phase. The dashed 
line is the result of Onsager's implicit solvent model.  See text.
\label{fig:Onsager}
}
\end{figure}
\paragraph{Dipole moments}
SMD implicit solvent calculations of the dipole moments of the fullerenediols
in solution were carried out at the spin-restricted 
\marginpar{\color{blue} SODS}
(same-orbitals-for-different-spins or SODS) level using geometries
optimized in the gas phase.  The simplest implicit solvent model is the 
Onsager model which we expect
to be a good first approximation for fullerendiols.  This is confirmed
in {\bf Fig.~\ref{fig:Onsager}} where we present a correlation plot between 
calculated gas-phase and aqueous-phase dipole moments.
The Onsager model predicts a straight line 
passing through the origin and with slope,
\begin{equation}
  m = 1 - \left( \frac{2 (\epsilon -1)}{2 \epsilon+1} \right)
  \left( \frac{n^2-1}{n^2+2} \right) \, . 
  \label{eq:SMD.1}
\end{equation}
Here $\epsilon$ is the dielectric constant of the surrounding water,
which is well-known to be close to 80, and $n$ is the refractive index of
solid C$_{60}$ which is 2.2 at 630 nm wavelength \cite{H91}.  Using these 
values gives a slope of $m=0.44906$ which corresponds to the dashed 
line in Fig.~\ref{fig:Onsager}.  We can conclude that the Onsager model is a 
reasonable first approximation to the SMD implicit solvent model in this case.

\begin{figure}
\begin{center}
\begin{tabular}{cc}
(a) & \\
(b) & \includegraphics[width=0.6\textwidth]{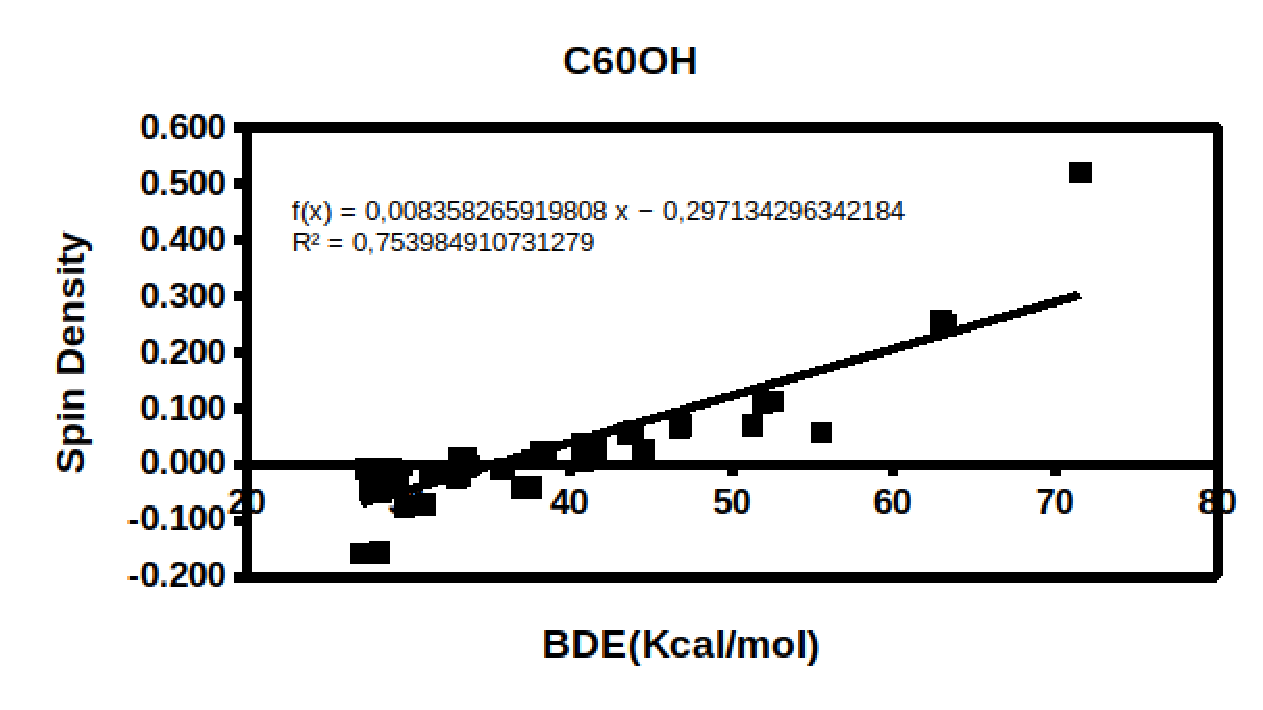} \\
(c) & \includegraphics[width=0.6\textwidth]{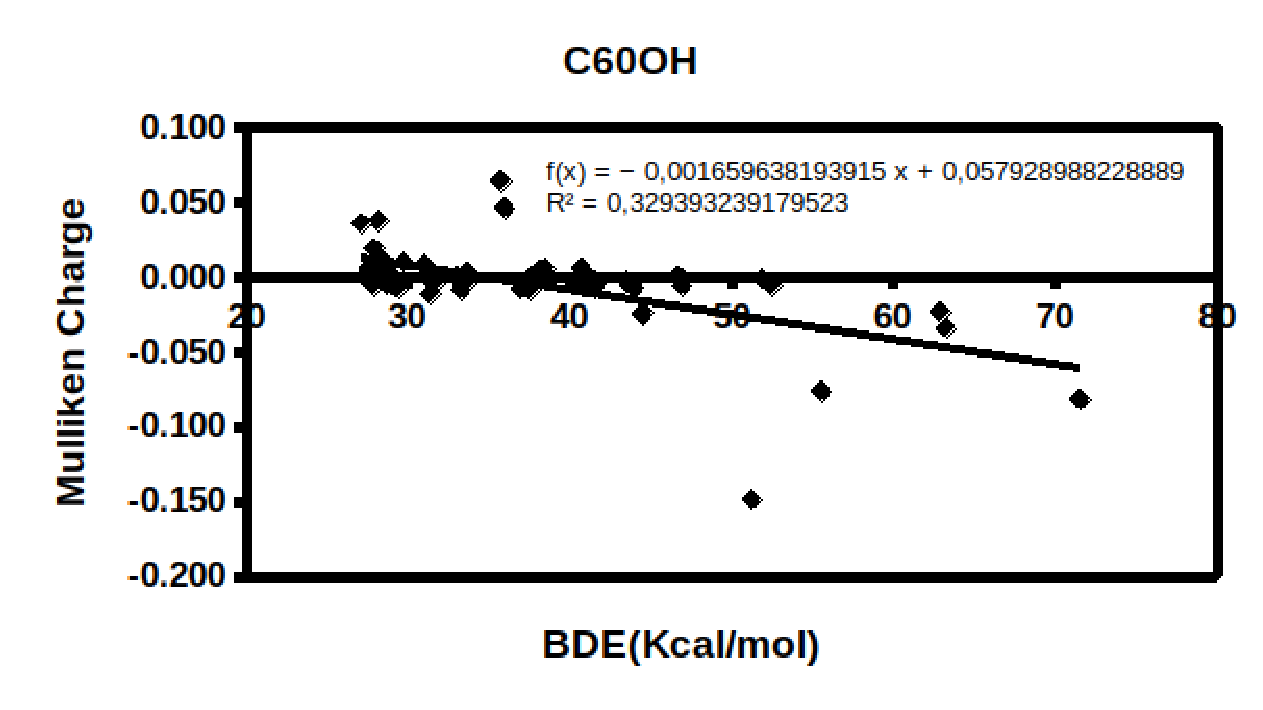} \\
    & \includegraphics[width=0.6\textwidth]{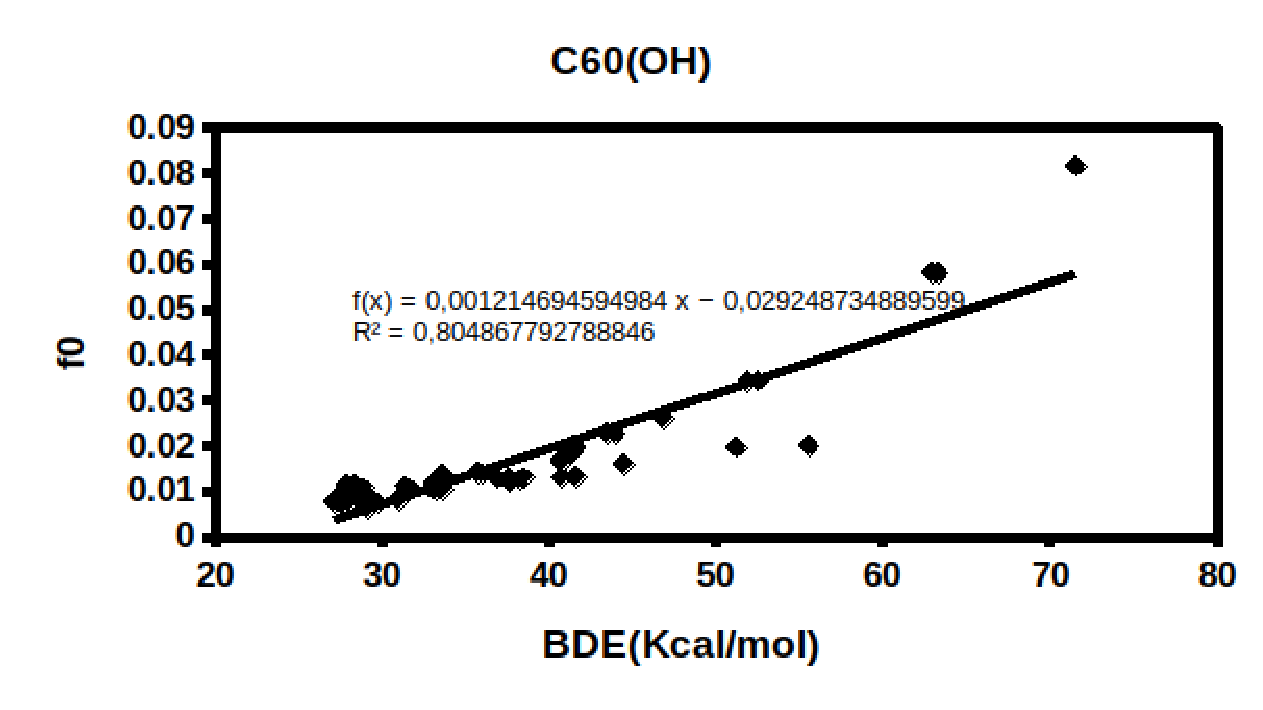} \\
\end{tabular}
\end{center}
\caption{
B3LYP(VWN3)/def2SVP/SMD reactivity indices calculated with 
an implicit solvent model.
\label{fig:SMDReactivityIndices}
}
\end{figure}

\paragraph{Reactivity indices}
Given such a large change in the dipole moments going between the gas-
and aqueous-phases, we might also except a large change in reactivity
indices.  As in Ref.~\cite{PEMC23}, we seek how well reactivity indices
\marginpar{\color{blue} BDE}
predict the C-O bond dissociation energy (BDE).  This latter is calculated
as the difference between the sum of the spin-unrestricted
\marginpar{\color{blue} DODS}
(different-orbital-for-different-spins or DODS) energies of 
$^\bullet$C$_{60}$(OH) + $^\bullet$OH and the SODS energy of C$_{60}$(OH)$_2$.
The gas-phase results confirmed that the BDE energy increases (i.e.,
the fullerenediol becomes more stabe) as the spin density of 
the reaction site $^\bullet$C$_{60}$(OH) increases.  Part (a) of
{\bf Fig.~\ref{fig:SMDReactivityIndices}} shows that this also holds
in our solution study.  The gas-phase results also confirmed the
electrophilic nature of the $^\bullet$OH radical because larger
BDEs were generally found at sites with more negative Mulliken charges.
Part (b) of Fig.~\ref{fig:SMDReactivityIndices} shows that this
continues to hold in aqueous solution.  Finally large values of the
radical Fukui function $f^0$ were also found to favor large values
of the BDE in the gas phase and part (c) of 
Fig.~\ref{fig:SMDReactivityIndices} shows that this
continues to hold in aqueous solution.
That the SMD aqueous solution results 
look very much like the corresponding gas-phase results 
\cite{PEMC23} is a nice confirmation that $^\bullet$OH remains
an electrophilic radical in solution even with respect to an ``electron
sponge'' such as C$_{60}$.

\begin{figure}
\begin{center}
\includegraphics[width=0.7\textwidth]{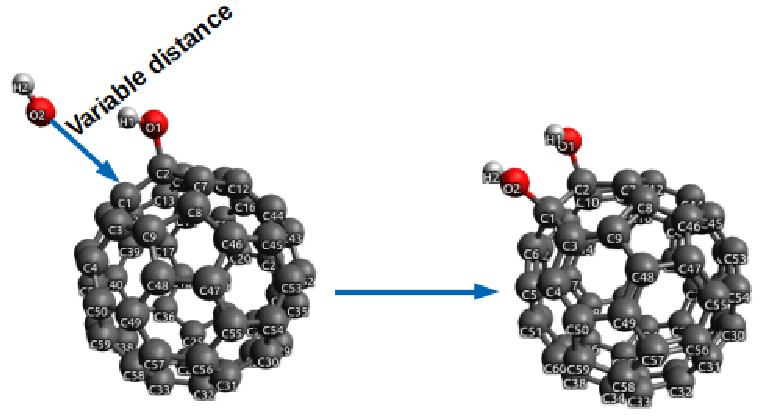} 
\end{center}
\caption{
Illustration for isomer {\bf 1} of a procedure used for isomers {\bf 1}, 
{\bf 2}, and {\bf 3}.  The vertical docking of $^\bullet$OH on 
$^\bullet$C$_{60}$OH was accomplished by begining from the fullerendiol
and using {\sc Molden} \cite{molden} to construct a Z-matrix.  Then only
the C-O distance was varied for carbon number 1, keeping all other bond
and dihedral angles fixed.  In particular the structures of $^\bullet$OH
and $^\bullet$C$_{60}$OH were kept fixed.
\label{fig:vertical}
}
\end{figure}

\paragraph{Exploratory study of PECs}
The large number of fullerenediol isomers (not to mention their various
conformers!) shown in Table~\ref{tab:isomercount}
represents a serious complication for studying the PECs for
reaction~\ref{eq:results.1}.  In order to keep things manageable, we have
chosen to only calculate PECs for for rigid vertical removal of 
$^\bullet$OH without any internal relaxation of either $^\bullet$C$_{60}$OH 
or $^\bullet$OH ({\bf Fig.~\ref{fig:vertical}}).  

\begin{figure}
\begin{center}
\includegraphics[width=0.9\textwidth]{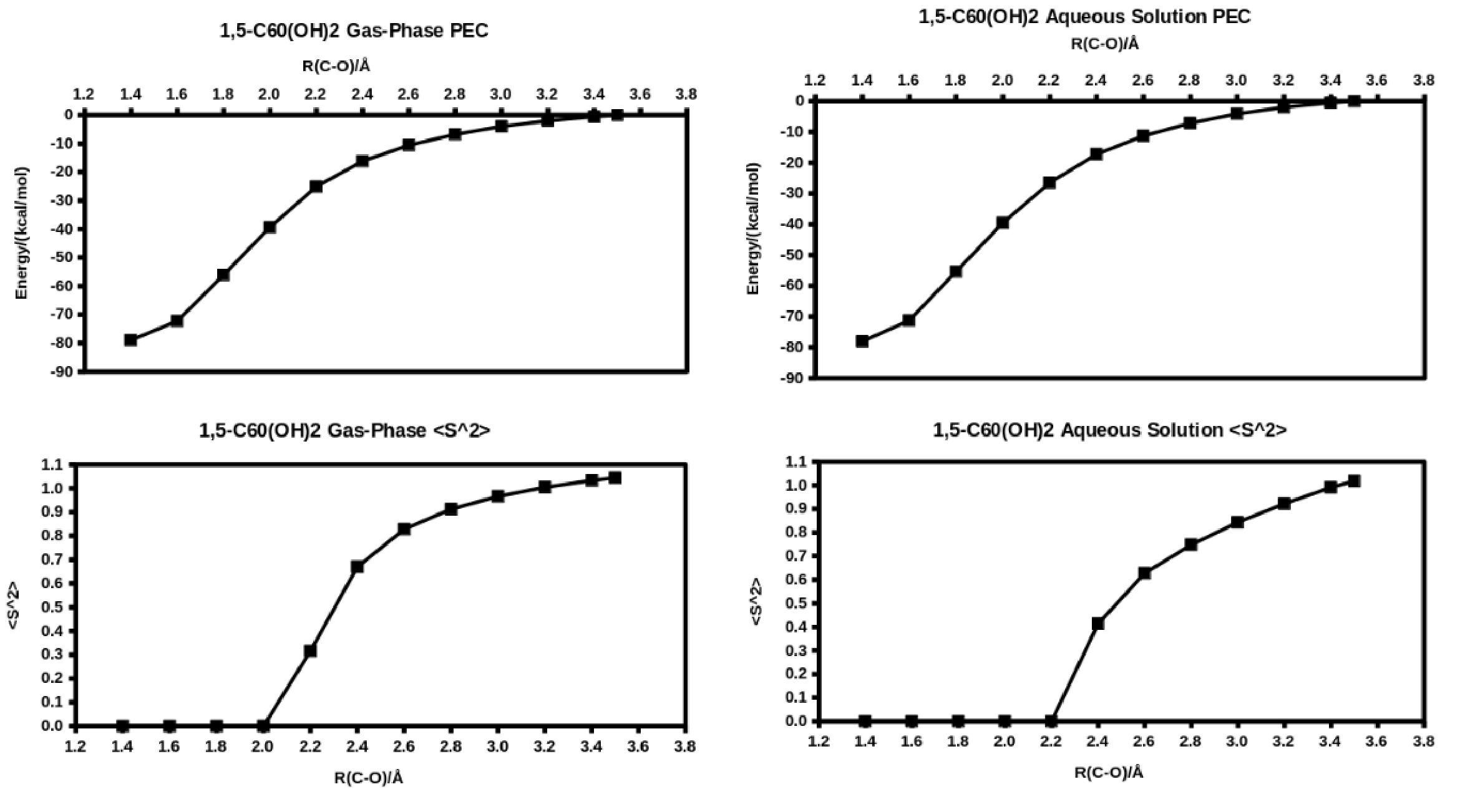} 
\end{center}
\caption{
1,5-C$_{60}$(OH)$_2$ (isomer {\bf 6}) vertical dissociation PECs 
and spin-contamination for the same geometries in gas phase and in 
aqueous solution.
\label{fig:PECandS2_5}
}
\end{figure}

Calculations were of the spin-unrestricted (DODS) type with symmetry 
breaking so as to be able to describe the breaking of the C-O single 
bond as correctly as possible.  Both energies and spin contamination 
($\langle \hat{S}^2 \rangle$) were monitored.  This produced much too
many results to given in the main body of the paper so we will just give
a few examples and some summarizing statistics.  Full results may be
\marginpar{\color{blue} SI}
found in the Supplementary Information (SI) associated with this article.
What we expected to see were curves such as those found for
1,5-C$_{60}$(OH)$_2$ and shown in {\bf Fig.~\ref{fig:PECandS2_5}}.
At large $R$(C-O), we expect to see a heavily spin-contaminated
open-shell singlet with $\langle \hat{S}^2 \rangle \approx 1$.
As the $^\bullet$C$_{60}$(OH) and $^\bullet$OH radicals approach
and form a bond, we expected a normal closed-shell singlet with
$\langle \hat{S}^2 \rangle = 0$.  Interestingly, this only happened
for about a quarter of the symmetry pairs studied.

\begin{figure}
\begin{center}
\includegraphics[width=0.8\textwidth]{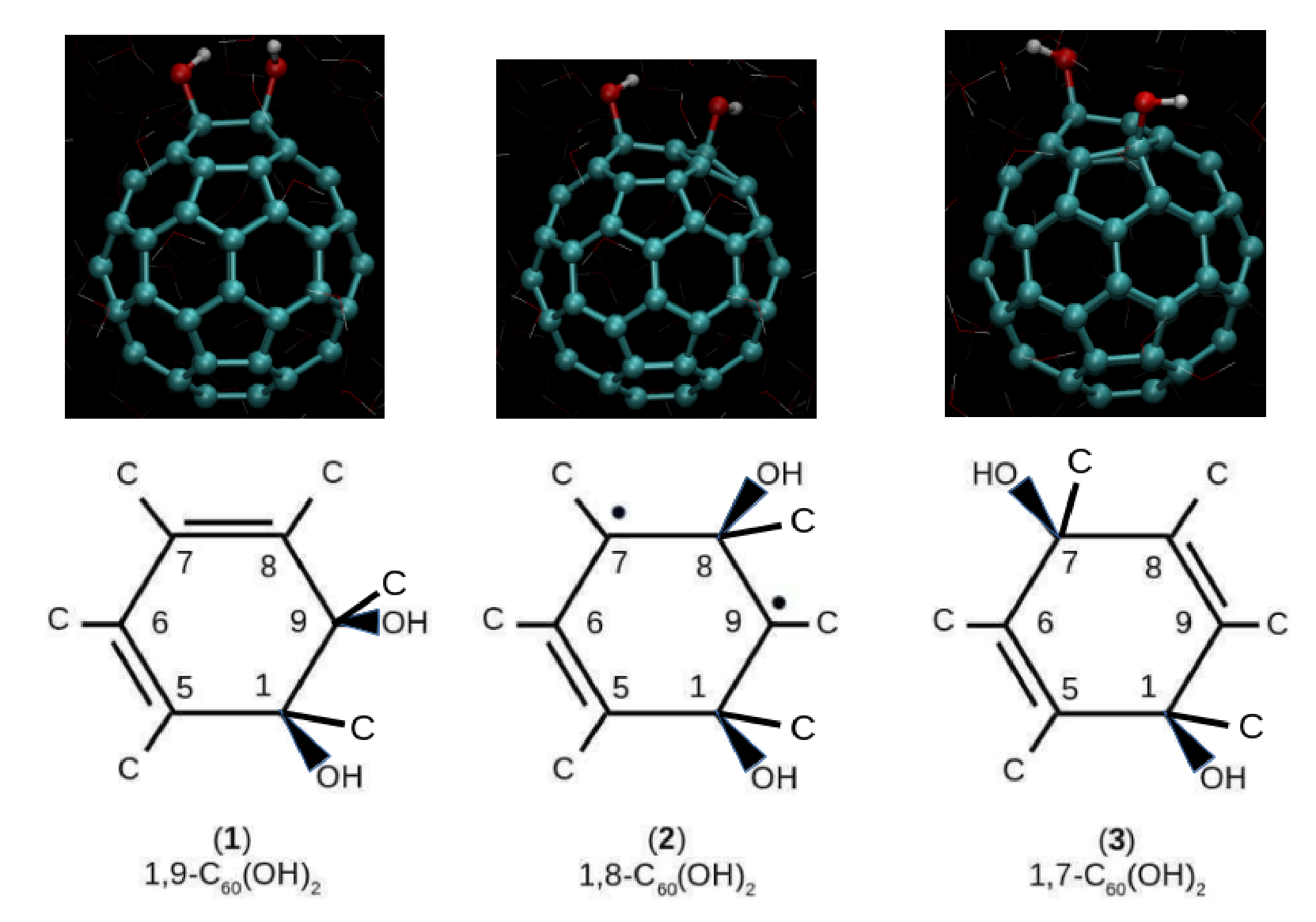}
\end{center}
\caption{
Full (top) and partial (bottom) fullerendiol structures showing 
the diradical nature of 1,8-C$_{60}$(OH)$_2$.  Note that there are
several other resonance structures that could be drawn for isomer {\bf 2}.
\label{fig:diradical}
}
\end{figure}

\begin{figure}
\begin{center}
\includegraphics[width=0.7\textwidth]{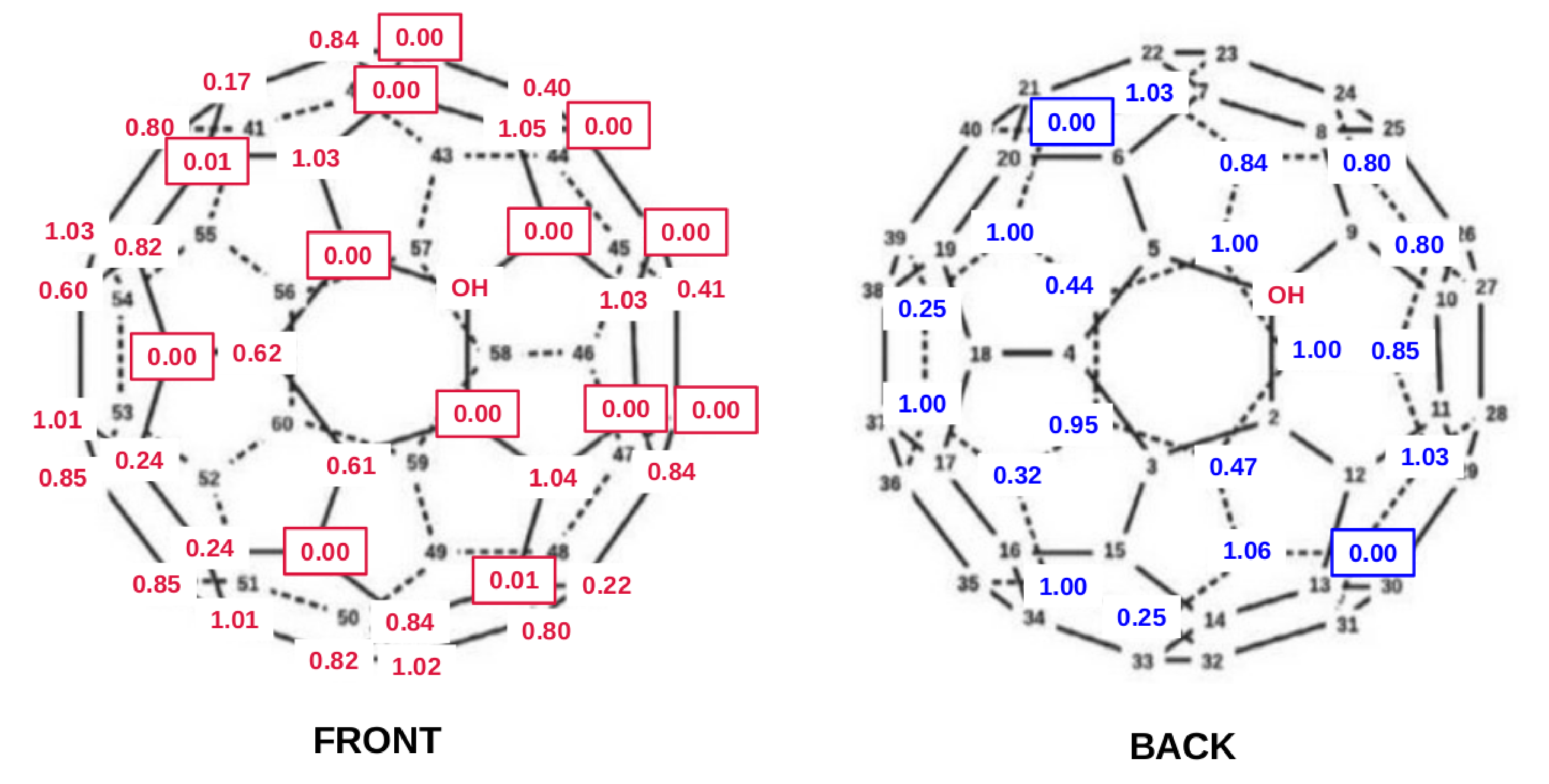} 
\end{center}
\caption{
Spin contamination: Averaged over gas-phase and aqueous-phase values,
the numbers show the degree of spin-contamination
at a C-O distance of 1.4 {\AA} for the {\em second} OH group.  Note
that the first OH group is always on carbon 1.  It is important to
note that the underlying figure is the same as in Fig.~\ref{fig:IUPAC}
because the carbon numbers are obscured by the red and blue numbers
showing the value of $\langle {\hat S}^2 \rangle$.
\label{fig:contaminationsites}
}
\end{figure}

Much like with the reactivity indices, the PECs and spin-contamination 
curves turn out to be remarkably similar in gas-phase and in aqueous solution.  
The degree of spin contamination for the different isomers is shown in 
{\bf Fig.~\ref{fig:contaminationsites}}.  It is apparent
that there are many fullerediols which are not simple closed-shell
singlets. At first thought, this may be a bit surprising as such structures
are rarely mentioned in the relevant literature.  On second thought,
it is easy to understand why radicaloid structures might be preferred
in many cases.  One such structure is illustrated in 
{\bf Fig.~\ref{fig:diradical}} and has been verified by examination of 
the spin density of 1,8-C$_{60}$(OH)$_2$.  In contrast, both 
1,7-C$_{60}$(OH)$_2$ and 1,9-C$_{60}$(OH)$_2$ are closed-shell singlets.

As expected, the numbers in Fig.~\ref{fig:contaminationsites} have near 
bilateral symmetry with respect to a mirror plane drawn through C1 and C9.  
Values of $\langle {\hat S}^2 \rangle$ vary from 0.00 to about 1.00 (in some 
cases we see $\langle {\hat S}^2 \rangle$ slightly exceeding 1.00).  
This later value is typical of what is expected for a diradical. 
Interestingly, the sites with zero spin contamination seem almost, but 
not quite, to alternate with sites with significant spin contamination.  
They are also mostly on the front side, rather than the back side, relative
to C1.
The value of $\langle {\hat S}^2 \rangle = 1$ has already been explained
\marginpar{\color{blue} LDS}
by the Lewis dot structure (LDS) shown in Fig.~\ref{fig:diradical}.  
The implications for further reactivity with a third $^\bullet$OH 
radical are evident.  However, we should remember that the unpaired spins may 
be widely distributed around the molecule in a way that corresponds to
multiple diverse resonance structures in the Lewis representation.
The reader is invited to explore which different LDSs may be drawn for
each isomer.

\begin{figure}
\begin{center}
\includegraphics[width=0.9\textwidth]{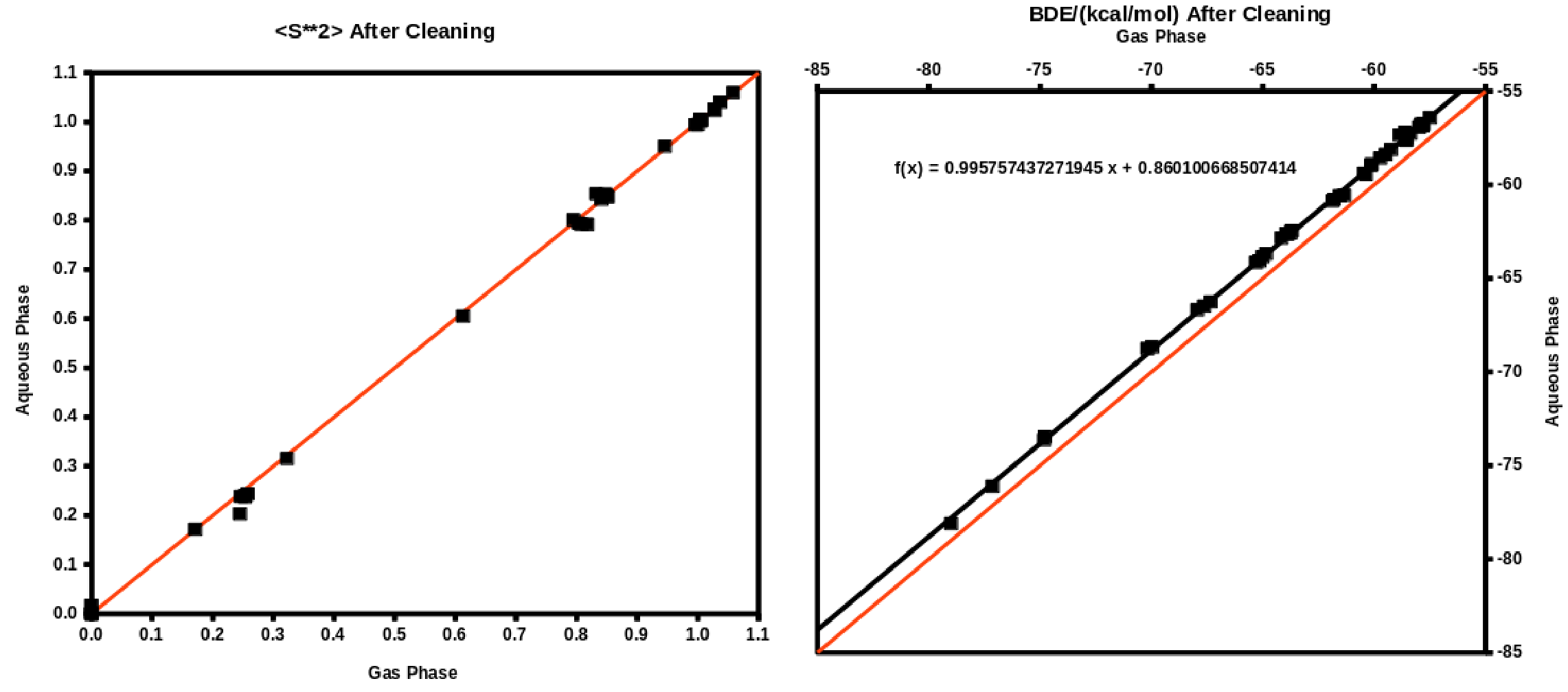} 
\end{center}
\caption{
Aqueous-phase versus gas-phase correlation graph of 
$\langle \hat{S}^2 \rangle$ and BDEs at $R$(C-O) = 1.4 {\AA}.
\label{fig:cleaned}
}
\end{figure}

Inspection of the "BDE" at $R$(C-O) = 1.4 {\AA} relative to a zero of 
energy at large $R$(C-O) for symmetry pairs showed that our results are 
plagued by more numerical problems than is the case for spin contamination.
Often these are in cases where spin contamination is high and there might
be more than one way to break symmetry.  When the BDEs at $R$(C-O) = 1.4 {\AA}
for symmetry pairs are very different, we also usually find significant
differences between the corresponding PECs and $\langle {\hat S}^2 \rangle$
curves (shown in the SI).  These numerical problems translate into an analysis
problem that we have only been able to solve by ``cleaning'' (i.e., removing
from further analysis) symmetry pairs whose BDEs at $R$(C-O) = 1.4 {\AA}
differ by more than 4 kcal/mol.  Specifically, all of the symmetry pairs
indicated by an asterisk in Table~\ref{tab:isomercount} have been removed
from the data.  The cleaned results for both $\langle {\hat S}^2 \rangle$
and the BDE at $R$(C-O) = 1.4 {\AA} are shown in {\bf Fig.~\ref{fig:cleaned}}.
It is seen that $\langle {\hat S}^2 \rangle$ is the same in gas phase
and in aqueous phase.  The gas-phase and aqueous-phase BDE are now linearly
related with the BDE being 0.86 kcal/mol higher in the gas phase than
in the aqueous phase.  This is consistent with the idea that the products, $^\bullet$C$_{60}$(OH)$_2$ and (especially) $^\bullet$OH are significantly stabilized relative to the reactant C$_{60}$(OH)$_2$ in aqueous solution.

\begin{figure}
\begin{center}
\includegraphics[width=0.9\textwidth]{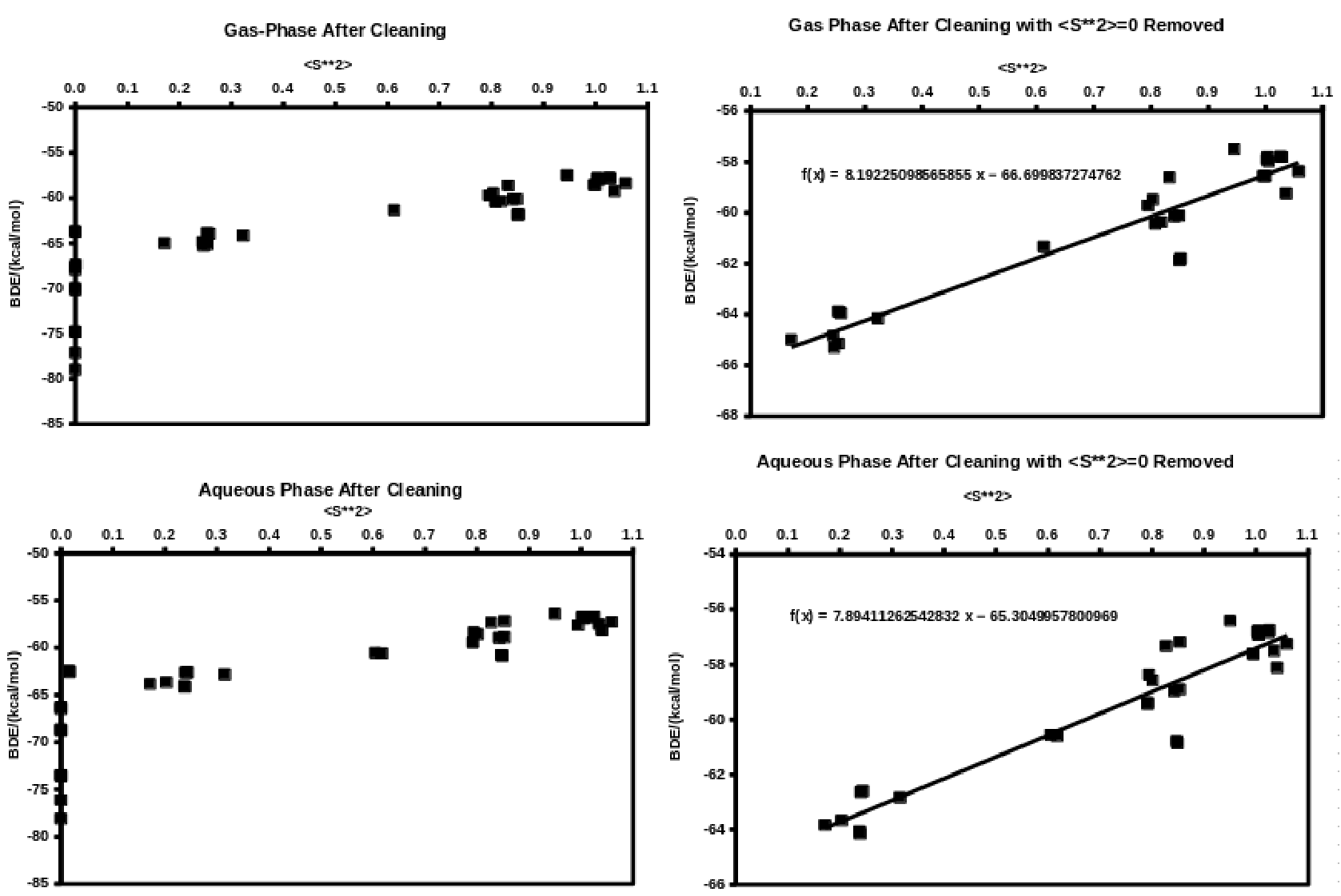} 
\end{center}
\caption{
BDEs as a function of $\langle \hat{S}^2 \rangle$ at $R$(C-O) = 1.4 {\AA}
after data cleaning.
\label{fig:S2BDEcorr}
}
\end{figure}

{\bf Figure~\ref{fig:S2BDEcorr}} provides a closer look at the correlation 
between the BDE and $\langle {\hat S}^2 \rangle$ in both gas phase 
and in aqueous solution.  For the isomers where 
$\langle {\hat S}^2 \rangle=0$, there is a spread
of BDEs governed by the chemical reactivity indices discussed above.
For $\langle {\hat S}^2 \rangle > 0$, there is a decrease in the BDE
as the spin-contimination increases.  The functional dependence
of the BDE as a function of $\langle {\hat S}^2 \rangle$ seems to be
essentially the same in gas and in aqueous solution, even if the BDE
in the gas phase is more highly bound by 0.86 kcal/mol.

\begin{figure}
\begin{center}
\includegraphics[width=0.9\textwidth]{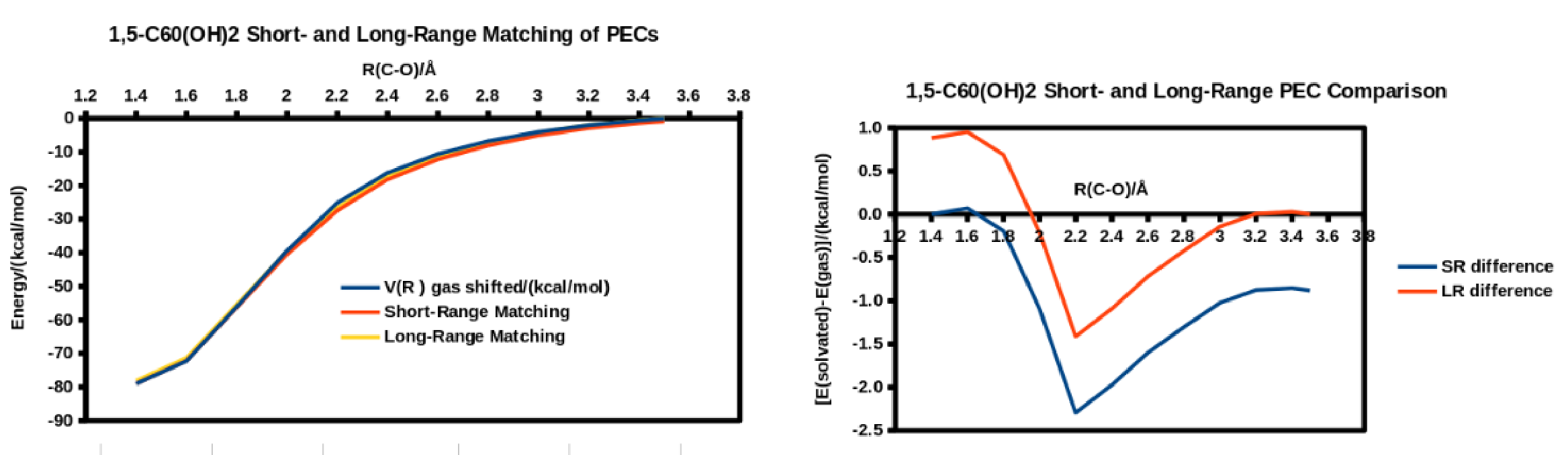} 
\end{center}
\caption{
Comparison of the short- and long-range shapes of the solution 
1,5-C$_{60}$(OH)$_2$ (isomer {\bf 6}) vertical dissociation PEC with 
the gas-phase PEC: left, PECs matched at short and long range;
right, corresponding solution minus gas phase PEC difference curves.
\label{fig:SRLR_5}
}
\end{figure}

By now it should be pretty clear that the PECs in gas-phase and in 
aqueous-phase for a given isomer are very similar.  This is illustrated
by {\bf Fig.~\ref{fig:SRLR_5}} and further illustrated by figures for
other isomers given in the SI.

\begin{figure}
\begin{center}
\includegraphics[width=0.9\textwidth]{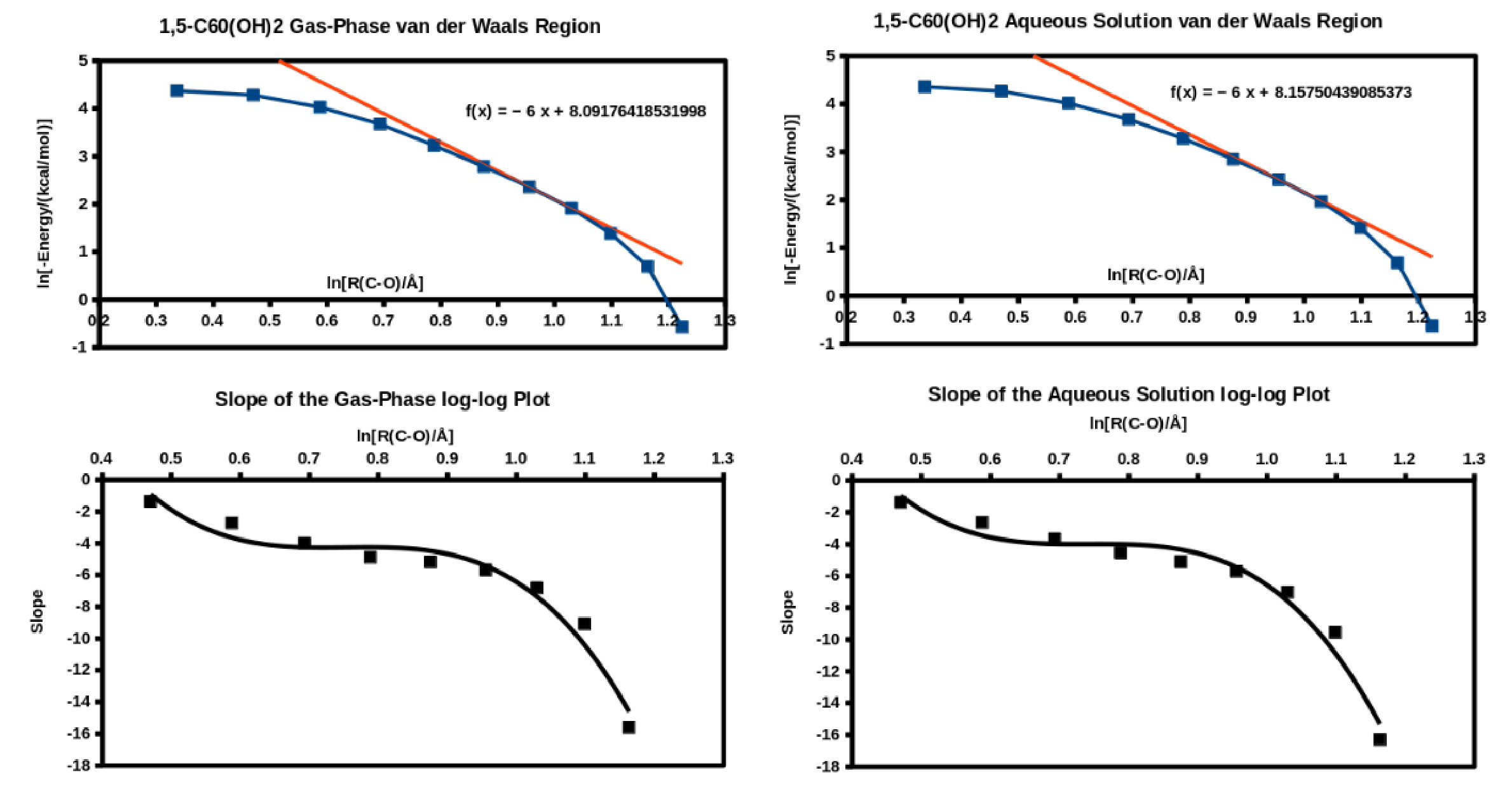} 
\end{center}
\caption{
Identification of the van der Waals region of the 1,5-C$_{60}$(OH)$_2$ 
(isomer {\bf 6}) vertical dissociation PECs. The upper row shows log-log 
plots of the PEC and tries to ``fit'' the van der Waals region. The lower 
row uses a finite difference method to calculate the derivative and then 
does a least-squares fit to a cubic polynomial.
\label{fig:vdW_5}
}
\end{figure}

Lastly, in our exploratory work, we wanted to identify at least some of
the responsible for the shape of the PECs.  We expected the long-distance 
behavior to go as $1/R^3$ which is typical of a dipole-dipole interaction.  
Instead we were surprised to see a region with a $1/R^6$ van der Waals 
interaction-like behavior ({\bf Fig.~\ref{fig:vdW_5}}).  There is a simple 
argument that DFT should be unable to describe the van der Waals $1/R^6$ 
part of the potential when the electron density of the separated fragments 
no longer overlap.  Our calculations do not contradict this idea as the $1/R^6$
is no longer present at very long distance.  However, this is a reminder
that a good functional can capture some of the $1/R^6$ behavior at 
moderately long distance.  Indeed one of the challenges when adding
van der Waals corrections to DFT is not to overcount any $1/R^6$ behavior
that may be already present in the functional.



%
%


\section{Concluding Discussion}
\label{sec:conclude}


Buckminsterfullerene (C$_{60}$) is such an excellent radical scavanger
that it has been called a ``radical sponge'' \cite{MML92}.  This makes
it interesting as a potential antioxidant for biological applications
provided that its solubility can be improved.  One way to improve the
solubility of C$_{60}$ in water is to add hydroxyl groups which may be
done in a variety of ways, including by successive addition of hydroxyl
radicals ($^\bullet$OH).  The relative stability of different fullerenols
($^\bullet$)C$_{60}$(OH)$_n$ (which is a radial for odd $n$) has been
studied theoretically in previous gas-phase work \cite{PEMC23}.  Whereas
the previous work treated additional fullerenols, the present work is 
limited to fullerendiols, C$_{60}$(OH)$_2$.  However the present work
does extend the previous work in three ways.

Firstly, the SMD implicit solvent model has been used to include the 
dielectric nature of water, albeit without including the effects of hydrogen
bonding.  As C$_{60}$(OH)$_2$ is nearly spherical, we expected that the
SMD model would give essentially the same result as the Onsager model
which is based upon a spherical cavity.  This was shown to be the case.
The effect of the implicit solvent model on the dipole moment of the
molecule is large and yet calculated reactivity indices show exactly the
same qualitative trends as in the gas phase, confirming that $^\bullet$OH 
is an electrophilic radical \cite{PEMC23}.  All of these calculations
were done using a spin-restricted formalism for C$_{60}$(OH)$_2$ and
a spin-unrestricted formalism for $^\bullet$C$_{60}$(OH) and for
$^\bullet$OH.

Secondly, we calculated potential energy curves (PECs) for the dissociation
reaction C$_{60}$(OH)$_2$ $\rightarrow$ $^\bullet$C$_{60}$(OH) + $^\bullet$OH
in both gas-phase and with the SMD implicit solvent model.  This required us 
to use a spin-unrestricted formalism throughout the reaction path so that
the spins could pair up in the same orbitals in C$_{60}$(OH)$_2$ but separate
into different orbitals in the products $^\bullet$C$_{60}$(OH) + $^\bullet$OH.
That still left us with a tremendous problem of the dissociation pathway of
59 C$_{60}$(OH)$_2$ isomers, each of which has multiple conformers.  We decided
to first carry out a crude vertical dissociation of each isomer which already
gave us some very interesting information.  In particular, the gas-phase and 
aqueous-phase PECs turn out to be quantitatively similar.  

Thirdly, we noticed that, contrary to our (and we suppose also that of other
workers in this field) expectations, it is not always true that 
$\langle \hat{S}^2 \rangle = 0$ for C$_{60}$(OH)$_2$.  In fact, it is only
true for about a quarter of the isomers.  We were able to convince ourselves 
by examining spin densities and by drawing resonance structures that 
1,8-C$_{60}$(OH)$_2$ has diradicaloid character and we expect that there must
also be di- or even pluriradicaloid character for many of the other 
fullerenediols.  This actually creates a new problem that we did not attempt
to address, namely that it is well established that there are several different
ways to break symmetry in magnetic clusters and finding the lowest energy 
broken energy solution is nontrivial.  What we did do was to recognize that
all but 3 of the fullerenediol isomers come in symmetry pairs and that these
almost always give quantitatively similar PECs and graphs of 
$\langle \hat{S}^2 \rangle$ along the dissociation pathway.  This allows us
to identify and discard pathways where the symmetry pairs behave too 
differently (which was the case for only 4 of the isomers).  None of the
PECs show any convincing evidence of a barrier along their PEC.

While we find it reassuring that many of our conclusions drawn from gas-phase
calculations are confirmed in our implicit solvent model, the most important
conclusion of the present work is the realization that arbitrary fullerenols
should and do have radicaloid character which should be taken into account
by symmetry breaking (as we have done) or by yet more sophisticated linear
combination of configuration state functions within some sort of wave function
or hybrid wave function/DFT approach.


\section*{Acknowledgements}
We gratefully acknowledge travel funding for AJE from the
U.S.-Africa Initative \cite{USAfrI}.
This work was partially funded by NIH R01GM108583 and R35GM151951 to GAC. 
Computing time  for this project was provided by the University of North 
Texas CASCaM  high-performance clusters NSF Grant Numbers CHE1531468 and 
OAC-2117247, by the NSF Extreme Science and Engineering Discovery Environment, 
ACCESS project Number TG-CHE160044, and the University of Texas at Dallas’ 
Cyberinfrastructure and Research Services.  AP would like to thank 
Pierre Girard for technical support in the context
of the Grenoble {\em Centre d'Experimentation du Calcul Intensif en Chimie}
({\em CECIC}) computers used for the {\sc Orca} calculations reported here.

\newpage


\begin{thebibliography}{10}
	
	\bibitem{MML92}
	C.~N. {McEwen}, R.~G. {McKay}, and B.~S. Larsen,
	\newblock {\sf \color{blue}{C$_{60}$} as a radical sponge},
	\newblock J. Am. Chem. Soc. {\bf 114}, 4412 (1992).
	
	\bibitem{C60skincream}
	{\sf \color{blue}The purest carbon 60 oil superantioxidant},
	\newblock https://purec60oliveoil.com/,
	\newblock Last accessed 5 March 2024.
	
	\bibitem{CMW+00}
	Z.~Chen, K.~Ma, G.~Wang, X.~Zhao, and A.~Tang,
	\newblock {\sf \color{blue}Structures and stabilities of {C$_{60}$(OH)$_4$} and
		{C$_{60}$(OH)$_6$} fullerenols},
	\newblock J. Mol. Struct. (Theochem) {\bf 498}, 227 (2000).
	
	\bibitem{RG04}
	J.~G. {Rodr\'{\i}guez-Zavala} and R.~A. {Guirado-L\'opez},
	\newblock {\sf \color{blue}Structure and energetics of polyhydroxylated carbon
		fullerenes},
	\newblock Phys. Rev. B {\bf 69}, 075411 (2004).
	
	\bibitem{KMT+08}
	K.~Kokubo, K.~Matsubayashi, H.~Tategaki, H.~Takada, and T.~Oshima,
	\newblock {\sf \color{blue}Facile synthesis of highly water-soluble fullerenes
		more than half-covered by hydroxyl groups},
	\newblock ACS Nano {\bf 2}, 327 (2008).
	
	\bibitem{FR09}
	E.~F. Fileti and R.~Rivelino,
	\newblock {\sf \color{blue}The {$^{13}$C NMR} properties of low hydroxylated
		fullerenes with density functional theory},
	\newblock Chem. Phys. Lett. {\bf 467}, 339 (2009).
	
	\bibitem{RTS+11}
	J.~G. {Rodr\'{\i}guez-Zavala}, F.~J. Tenorio, C.~Samaniego, C.~I.
	{M\'endez-Barrientos}, F.~G. {Pen\~na-Lecona}, J.~{Mu\~noz-Maciel}, and
	R.~{Flores-Moreno},
	\newblock {\sf \color{blue}Theoretical study on the sequential hydroxylation of
		{C$_{82}$} fullerene based on {F}ukui function},
	\newblock Mol. Phys. {\bf 109}, 1771 (2011).
	
	\bibitem{UYA+14}
	H.~Ueno, S.~Yakamura, R.~S. Arastoo, T.~Oshima, and K.~Kokubo,
	\newblock {\sf \color{blue}Systematic evaluation and mechanistic investigation
		of antioxidant activity of fullerenols using $\beta$-carotene bleaching
		assay},
	\newblock J. Nanomater. {\bf 2014}, 802596 (2014).
	
	\bibitem{SCP+16}
	K.~N. Semenov, N.~A. Charykov, V.~N. Postnov, V.~V. Sharoyko, I.~V.
	Vorotyntsev, M.~M. Galaguzda, and I.~V. Murin,
	\newblock {\sf \color{blue}Fullerenols: {P}hysiochemical properties and
		applications},
	\newblock Prog. Solid State Chem. {\bf 44}, 59 (2016).
	
	\bibitem{AKMM17}
	S.~Afreen, K.~Kokubo, K.~Muthoosamy, and S.~Manickam,
	\newblock {\sf \color{blue}Hydration or hydroxylation: direct synthesis of
		fulleronol from pristine fullerene {[C$_{60}$]} {\em via} acoustic caviation
		in the presence of hydrogen peroxide},
	\newblock RSC Adv. {\bf 7}, 31930 (2017).
	
	\bibitem{KT17}
	S.~Keshri and B.~L. Tembe,
	\newblock {\sf \color{blue}Thermodynamics of hydration of fullerols
		{[C$_{60}$(OH)$_n$]} and hydrogen bond dynamics in their hydration shells},
	\newblock J. Chem. Phys. {\bf 146}, 074501 (2017).
	
	\bibitem{VGG18}
	M.~V. {Velarde-Salcedo}, M.~Gallo, and R.~A. {Guirado-L\'opez},
	\newblock {\sf \color{blue}Low hydroxylated fullerenes: {S}tability, thermal
		behavior, and vibrational properties},
	\newblock J. Phys. Chem. {\bf 122}, 13117 (2018).
	
	\bibitem{WGZ18}
	Z.~Wang, Z.~Gao, and Y.~Zhao,
	\newblock {\sf \color{blue}Mechanisms of antioxidant activities of fullerenols
		from first-principles calculation},
	\newblock J. Phys. Chem. A {\bf 122}, 8183 (2018).
	
	\bibitem{KSV+19}
	E.~S. Kovel, A.~A. Sachkova, N.~G. Vnukova, G.~N. Churilov, E.~M. Knyazeva, and
	N.~S. Kudryasheva,
	\newblock {\sf \color{blue}Antioxidant activity and toxicity of fullerenols via
		bioluminescence signaling: {R}ole of oxygen substituents},
	\newblock Int. J. Mol. Sci. {\bf 20}, 2324 (2019).
	
	\bibitem{ZSM+20}
	P.~Zygouri, K.~Spyrou, E.~Mitsari, M.~Barrio, , R.~Macovez, M.~Patila,
	H.~Stamatis, I.~I. Verginadis, A.~P. Velalopoulou, A.~M. Evangelou,
	Z.~Sideratou, D.~Gournis, and P.~Rudolf,
	\newblock {\sf \color{blue}A facile approach to hydrophilic oxidized fullerenes
		and their derivatives as cytotoxic agents and supports for nanobiocatalytic
		systems},
	\newblock Scientific Reports {\bf 10}, 8244 (2020).
	
	\bibitem{PEMC23}
	A.~Ponra, A.~J. Etindele, O.~Motapon, and M.~E. Casida,
	\newblock {\sf \color{blue}Binding energies for successive addition reaction of
		{$^\bullet$OH} with {C$_{60}$}: {A} laboratory for testing frontier molecular
		orbital theory},
	\newblock Adv. Quant. Chem. {\bf 8}, 2023 (2023).
	
	\bibitem{MCC+24}
	A.~{Moreno-Ceballos}, M.~E. Castro, N.~A. Caballero, L.~Mammino, and F.~J.
	Melendez,
	\newblock {\sf \color{blue}Implicit and explicit solvent effects on the global
		reactivity and the density topological parameters of the preferred conformers
		of caespitate},
	\newblock Computation {\bf 12}, 5 (2024).
	
	\bibitem{FS19}
	P.~P. Feh\'er and A.~Stirling,
	\newblock {\sf \color{blue}Assessment of reactivities with explicit and
		implicit solvent models: {QM/MM} and gas-phase evaluation of three different
		{Ag}-catalysed furan ring formation routes},
	\newblock New J. Chem. {\bf 43}, 15706 (2019).
	
	\bibitem{PEMC21}
	A.~Ponra, A.~J. Etindele, O.~Motapon, and M.~E. Casida,
	\newblock {\sf \color{blue}Practical treatment of singlet oxygen with
		density-functional theory and the multiplet-sum method},
	\newblock Theo. Chem. Acc. {\bf 140}, 154 (2021).
	
	\bibitem{PBEC23}
	A.~Ponra, C.~Bakasa, A.~J. Etindele, and M.~E. Casida,
	\newblock {\sf \color{blue}Diagrammatic multiplet-sum method ({MSM})
		density-functional theory ({DFT}): {I}nvestigation of the transferability of
		integrals in ``simple'' {DFT}-based approaches to multi-determinantal
		problems},
	\newblock J. Chem. Phys. {\bf 159}, 244306 (2023).
	
	\bibitem{MCT09}
	A.~V. Marenich, C.~J. Cramer, and D.~G. Truhlar,
	\newblock {\sf \color{blue}Universal solvation model based on solute electron
		density and on a continuum model of the solvent defined by the bulk
		dielectric constant and atomic surface tensions},
	\newblock J. Phys. Chem. B {\bf 113}, 6378 (2009).
	
	\bibitem{O36}
	L.~Onsager,
	\newblock {\sf \color{blue}Electric moments of molecules in liquids},
	\newblock J. Am. Chem. Soc. {\bf 58}, 1486 (1936).
	
	\bibitem{gaussian}
	M.~ea~Frisch,
	\newblock {\sf \color{blue}Gaussian 16 revision a. 03}, 2016.
	
	\bibitem{B73}
	C.~J.~F. B\"ottcher,
	\newblock {\em Theory of Electric Polarization, Vol. I, Dielectrics in Static
		Fields},
	\newblock Elsevier Scientific Publishing Company, New York, 1973.
	
	\bibitem{P63}
	E.~M. Purcell,
	\newblock {\em Electricity and Magnetism}, volume~2 of {\em Berkeley Physics
		Series},
	\newblock McGraw-Hill, New York, 1963.
	
	\bibitem{H91}
	D.~Huffman,
	\newblock {\sf \color{blue}Solid {C$_{60}$}},
	\newblock Physics Today {\bf 44}, 22 (1991).
	
	\bibitem{PCM+02}
	W.~M. Powell, F.~Cozzi, G.~P. Moss, C.~Thilgen, R.~J. Hwu, and A.~Yerin,
	\newblock {\sf \color{blue}Nomenclature for the {C$_{60}-I_h$} and
		{C$_{70}-D_{5h(6)}$} fullerenes ({IUPAC} recommendations (2002))},
	\newblock Pure Appl. Chem. {\bf 74}, 629 (2002).
	
	\bibitem{N18}
	F.~Neese,
	\newblock {\sf \color{blue}Software update: {T}he {\sc orca} program system,
		version 4.0},
	\newblock Wiley Interdisciplinary Reviews: Computational Molecular Science {\bf
		8}, e1327 (2018).
	
	\bibitem{B93}
	A.~D. Becke,
	\newblock {\sf \color{blue}Density-functional thermochemistry. {III}. {T}he
		role of exact exchange},
	\newblock J. Chem. Phys. {\bf 98}, 5648 (1993).
	
	\bibitem{LYP88}
	C.~Lee, W.~Yang, and R.~G. Parr,
	\newblock {\sf \color{blue}Development of the colle-salvetti correlation-energy
		formula into a functional of the electron density},
	\newblock Phys. Rev. B {\bf 37}, 785 (1988).
	
	\bibitem{WA05}
	F.~Weigend and R.~Ahlrichs,
	\newblock {\sf \color{blue}Balanced basis sets of split valence, triple zeta
		valence and quadruple zeta valence quality for {H} to {Rn}: {D}esign and
		assessment of accuracy},
	\newblock Phys. Chem. Chem. Phys. {\bf 7}, 3297 (2005).
	
	\bibitem{MDC24}
	D.~Magero, A.~M. H.~M. Dargouth, and M.~Casida,
	\newblock {\sf \color{blue}Test of the orbital-based {LI3} index as a predictor
		of the height of the {$^3$MLCT $\rightarrow$ $^3$MC} transition-state barrier
		for gas-phase {[Ru(N$\wedge$N)$_3$]$^{2+}$}},
	\newblock J. Photochem. Photobio. A: Chem. {\bf 451}, 115502 (2024).
	
	\bibitem{DMC24}
	A.~M. H.~M. Dargouth, D.~Magero, and M.~E. Casida,
	\newblock {\sf \color{blue}Test of the orbital-based {LI3} index as a predictor
		of the height of the {$^3$MLCT $\rightarrow$ $^3$MC} transition-state barrier
		for {[Ru(N$\wedge$N)$_3$]$^{2+}$} polypyridine complexes in {CH$_3$CN}},
	\newblock J. Phys. Chem. A {\bf XXX}, yyyyyy (2024?).
	
	\bibitem{molden}
	B.~Schaftenaar,
	\newblock {\sf \color{blue}{\sc MOLDEN}: {A} pre- and post processing program
		of molecular electronic structure},
	\newblock https://www3.cmbi.umcn.nl/molden/,
	\newblock Last accessed 22 May 2021.
	
	\bibitem{USAfrI}
	{\sf \color{blue}{U.S.-Africa Initative}},
	\newblock https://usafricainitative.org,
	\newblock Last accessed 20 February 2020.
	
\end{thebibliography}
\end{document}